# ODRŽAVANJE RAČUNARSKIH SISTEMA

# MAINTENANCE OF COMPUTER SYSTEMS

**Samir Lemeš, doc.dr.**
Univerzitet u Zenici, Mašinski fakultet
Zenica, Bosna i Hercegovina

**REZIME**
*Računarski hardware i software su već duže vrijeme resursi bez kojih je nezamislivo savremeno poslovanje bilo koje organizacije, od proizvodne do uslužne. Održavanje računarskih sistema je aspekt poslovanja kojem se ne poklanja dovoljno pažnje. U ovom radu date su neke preporuke za izbor pristupa održavanju računarskih sistema, zasnovane na dugogodišnjem iskustvu održavanja ovih sistema na Univerzitetu u Zenici.*

**Ključne riječi:** održavanje, software, hardware, računarski sistemi

**ABSTRACT**
*Computer hardware and software are resources without which the modern business of any organization, from manufacturing to services, is impossible. Not enough attention is being payed to maintenance of computer systems as an aspect of business. This paper gives some recommendations for the selection of the computer systems maintenance approach, based on many years of experience maintaining these systems at the University of Zenica.*

**Key words:** maintenance, software, hardware, computer systems

## 1. UVOD

Informatizacija poslovanja savremenih organizacija je dovela do velika zavisnosti čitavih sistema od računarske podrške. Ta podrška se može posmatrati odvojeno, kao hardware, software, komunikacije, sigurnost, licenciranje, ali su svi ti aspekti međusobno toliko povezani, da se ne mogu posmatrati izolovano, nego samo kao dio cjeline. Pored međusobne povezanosti, ti sistemi se isprepliću i s drugim aspektima poslovanja, tako da je za izbor strategije održavanja tog sistema potrebno imati složen i sveobuhvatan skup informacija. Posebnu težinu tom problemu daje dinamičnost računarskih sistema, koji se mijenjaju tako brzo da to nekad predstavlja ogroman finansijski teret, kako za mala i srednja preduzeća, tako i za velike poslovne sisteme.

Pristup održavanju računarskih sistema najčešće zavisi od veličine organizacije. Održavanje računarskih sistema se povjerava nekom od zaposlenih, koji je najstručniji za tu oblast, koriste se usluge treće strane, odnosno specijaliziranih informatičkih kompanija, a samo veliki sistemi mogu priuštiti posebnu informatičku službu.

## 2. PRETHODNA ISTRAŽIVANJA

Brojni autori su istraživali ovaj problem s različitih aspekata.

Problemi održavanja računarskih sistema su na samom početku informatizacije poslovanja bili posmatrani većinom kao problemi ograničeni na hardware. Schneider je u [1] dao pregled osnovnih operacija koje prosječan korisnik s ograničenim poznavanjem hardware-a može da obavlja kako bi obezbijedio nesmetano funkcionisanje računarske opreme, u njegovom slučaju mainframe računara. U organizacijama kao što su univerziteti ili istraživački centri se podrazumijeva da osoblje ima dovoljno vještina i informatičke pismenosti da jednostavne kvarove može identifikovati i otkloniti. Pristup preventivnog održavanja je tad podrazumijevao obezbjeđenje ambijentalnih uslova, temperature, stabilnog izvora napajanja električnom energijom, zaštite od statičkog elektriciteta, vlage i sl.

Shui-Shun et al. su u [2] analizirali problem održavanja notebook računara putem internet kolaborativne platforme. Na osnovu povratnih informacija prikupljenih od korisnika, razvili su informacioni sistem koji prati svaki računar koji je vraćen dobavljaču u garantnom roku ili koji se nalazi u postupku servisiranja. Na taj način smanjene su potrebe za rezervnim dijelovima, njihovo skladištenje je optimizirano, i skraćeno je vrijeme isporuke rezervnih dijelova jer se ta nabavka vrši putem mreže ovlaštenih servisa, a informacioni sistem bira servis koji je fizički najbliži korisniku.

Brinkman i Roubieu su u [3] dali diskusiju o vrstama podataka o kojima treba voditi evidenciju i kako se te informacije mogu primijeniti na stalno planiranje i donošenje odluka o upravljanju i održavanju računarske opreme. Oni su ukazali na ogroman značaj vođenja dokumentacije i evidencije o računarskim sistemima i njihovim komponentama, bez koje nema pravilnog održavanja.

Ward se u [4] kritički osvrnula na korištenje specijaliziranih servisa za održavanje u Velikoj Britaniji, te ukazala na njihovu fleksibilnost sa aspekta troškova. Kans je u [5] dao pregled sistema za održavanje i upravljanje IT (*Maintenance management information technology – MMIT*) u posljednjih 40 godina i njihovo poređenje s drugim IT sistemima. Zaključio je da se fokus MMIT promijenio u nekoliko aspekata tokom posmatranog perioda, od tehnologije ka korištenju, od funkcije održavanja do održavanja kao integralnog dijela poslovanja, od podrške reaktivnom održavanju do proaktivnog održavanja, te od operativnog do strateškog pristupa održavanju. Napredak u MMIT je općenito pratio razvoj informacionih tehnologija.

Tie et al. su u [6] prikazali problem održavanja računarskih centara u univerzitetskim laboratorijima. Na osnovu statističkih pokazatelja, stanje računarske opreme su podijelili na četiri moguća statusa: normalno stanje, zabrinjavajuće stanje, stanje upozorenja i stanje otkaza. Na osnovu tih stanja, predložili su i različite metode održavanja, čime su značajno smanjili potrebni nivo održavanja.

Yu-Ting i Ambler u [7] su razvili model za simulaciju održavanja personalnih računara, na osnovu kojeg se potrebe za održavanjem minimiziraju u ranoj fazi razvoja proizvoda. Njihov pristup podrazumijeva algoritam za donošenje odluka na osnovu više utjecajnih faktora, koji optimizira komponente sistema s ciljem minimiziranja troškova održavanja.

Zhang i Wang su se u [8] osvrnuli na okolišni aspekt održavanja, analizirajući koliki je utjecaj na okoliš računarskih sistema, uglavnom sa aspekta potrošnje energije, ali i sa aspekta zbrinjavanja elektronskog otpada, jer se radni vijek računarske opreme smanjuje, čime se povećava opterećenje na okoliš i raste potreba za reciklažom odbačene računarske opreme.

Barak i Malik su u [9] pokušali analizirati integrirani pristup održavanju i hardware-a i software-a. Dali su matematički model za srednje vrijeme do otkaza sistema (*Mean Time to System Failure – MTSF*) i na osnovu njega izvršili analizu troškova održavanja. Utvrdili su da pouzdanost sistema raste s rastom prediktivnog održavanja, ali se smanjuje s produženjem vremena rada sistema.

## 3. ODRŽAVANJE HARDWARE-A

O održavanju elektronskih komponenti računarskih sistema treba voditi računa od najranije faze informatizacije poslovanja, jer se početne uštede poslije mogu pokazati kao veliko opterećenje, kako finansijsko, tako i operativno. Na primjer, investicija u UPS (neprekidni izvor napajanja) je simbolična, a značajno smanjuje šanse pojave kvarova hardware-a. Neki dobavljači čak uslovljavaju garanciju za opremu napajanjem preko UPS-a.

### 3.1. Nabavka

Nekad su kod nabavke računara jedina garancija bili takozvani *brand-name* računari, koji su višom cijenom garantovali i pouzdaniji rad, jer su kombinacije komponenti višestruko rigorozno testirane od strane proizvođača. Performanse *no-name* i *brand-name* sistema se sve manje razlikuju (slike 1 i 2), a i cijene se mogu kretati od raspona 1:1 do 1:3. Puno važniji aspekt od pouzdanosti kod izbora konfiguracije je u stvari namjena, odnosno software koji će se koristiti. Vjerovatno će *brand-name* računar raditi bez ozbiljnijih kvarova i duže od 10 godina, ali će se za isti taj period promijeniti nekoliko generacija operativnih sistema i aplikativnog software-a, čime takva dugovječnost opreme postaje suvišna i ne opravdava visoku nabavnu cijenu.

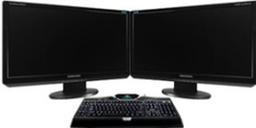
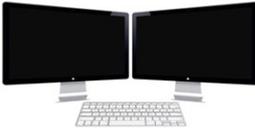
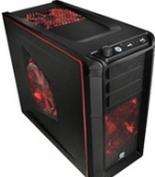
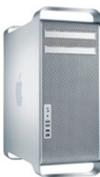
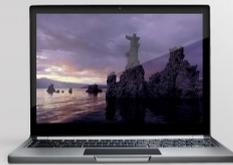
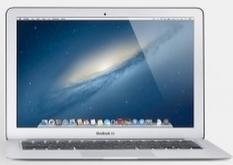

*Slika 1. Poređenje Windows PC i Mac radne stanice (izvor: http://www.techglint.com, jan. 2012)*

*Slika 2. Google Chromebook Pixel i MacBook Air (izvor: http://www.gizmag.com, feb. 2013)*

### 3.2. Dijagnostika

Osnovnu dijagnostiku računarskih sistema često mogu vršiti i sami korisnici, jer se popravke hardware-a svode na zamjenu komponenti. Nije teško utvrditi da li je uzrok kvara, npr. neispravan monitor, tastatura ili ventilator na procesoru. Nekad je teže odrediti da li se radi o software-skom (neispravan *driver*) ili hardware-skom kvaru, nego koja je komponenta u kvaru. Na slici 3 je prikazan veoma čest problem koji dovodi do "smrzavanja" računara, jer se zbog nakupljene prašine nedovoljno hladi CPU. Na slici 4 je prikazana poruka operativnog sistema o prekidu mrežne konekcije, koja na prvi pogled ukazuje na to da je konektor mrežnog kabla isključen, ali ista poruka će se pojaviti i ako neki od mrežnih uređaja (*switch*) ostane bez napajanja.

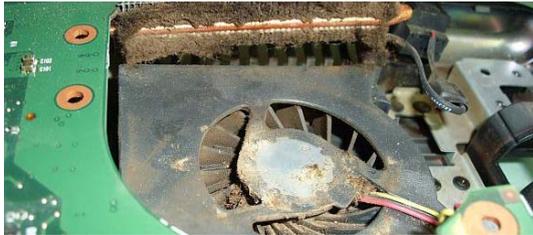 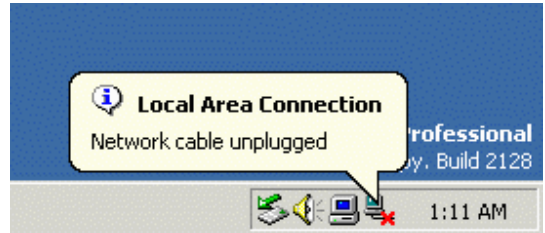

*Slika 3. Jedan od najčešćih problema u radu računara je zaprljan ventilator za hlađenje CPU (izvor: http://www.daftdog.co.uk, feb. 2013)*

*Slika 4. Indikacija problema s mrežnom konekcijom kod koje sama poruka operativnog sistema ne daje dovoljno informacija*

### 3.3. Otklanjanje kvarova

Zamjenu nekih komponenti (tastatura, miš) može izvršiti svaki korisnik računara. Prilikom zamjene komponenti koje zahtijevaju otvaranje kućišta računara treba voditi računa o kršenju uvjeta garancije, jer je to najčešći uzrok nepriznavanja odredbi garantnog roka od strane isporučioca opreme. Prilikom servisiranja izvan prostorija korisnika, potrebno je voditi računa o sigurnosti informacija, jer servis ima neograničen pristup podacima na računaru na kojem vrši popravke. Taj se problem rješava ili potpisivanjem izjave o povjerljivosti ili servisiranjem na licu mjesta, uz stalni nadzor ovlaštenog osoblja.

### 3.4. Evidencija

Veoma je važno voditi ispravnu evidenciju računarske opreme. Prema važećim propisima, obavezno je vođenje liste stalnih sredstava, koja skoro nikad ne opisuje dovoljno detalja za računarsku opremu. Na primjer, USB memorijski stik se vodi kao stalno sredstvo umjesto kao sitni inventar. Drugi primjer je evidentiranje računara bez podataka o konfiguraciji (kapacitet HDD, RAM,...), što otvara mogućnost da osoblje otuđi dio opreme, a da o tome ne postoji evidencija. Evidencija se treba koristiti i za planiranje nabavke i obnavljanja opreme.

### 3.5. Zbrinjavanje elektronskog otpada

Ovom aspektu održavanja hardware-a u BiH se još ne poklanja potrebna pažnja i otpisani i neispravni računari i njihove komponente često završavaju na komunalnim deponijama, jer ne postoji organizovani i zakonom uređeni sistem za njihovo zbrinjavanje i reciklažu.

### 4. ODRŽAVANJE SOFTWARE-A
### 4.1. Rezervne kopije podataka

Jedini pouzdan način da se obezbijedi sigurnost digitalno pohranjenih informacija je izrada rezervnih kopija podataka. Obaveza izrade rezervnih kopija se može uvesti kroz sistem upravljanja kvalitetom. Za izradu rezervnih kopija koriste se prenosivi mediji (USB *flash drive*, CD-RW, DVD-RW), te besplatni ili komercijalni servisi za pohranjivanje podataka na internetu (*Dropbox*, *Google Drive*, *Microsoft Live*,...). Na serverima je uobičajeno koristiti RAID tehnologiju koja redudantnošću pohrane podataka obezbjeđuje automatske rezervne kopije. U posljednje vrijeme se RAID koristi čak i na desktop računarima, bilo da je implementiran kroz hardware (disk kontroler) ili software (operativni sistem).

### 4.2. Licenciranje

Sve više proizvođača software-a prelazi na dinamičke licence za svoje proizvode. Umjesto da se kupovinom licence postane vlasnik dozvole za korištenje proizvoda, proizvođači nude različite opcije vremenski ograničenih prava korištenja. Na taj način obavezuju korisnika da vrši redovno plaćanje na duži vremenski period, a zauzvrat se nudi automatsko ažuriranje novih verzija i podrška korisnicima. Često se za podršku korisnicima koriste tehnike udaljenog pristupa, kao što su *Windows Remote Desktop Connection* ili *TeamViewer*.

### 4.3. Ažuriranje verzija

Osnovni problem kod korištenja aplikativnog software-a je prelazak na novu verziju kad se ona pojavi. Nekad se nova verzija toliko razlikuje da je potrebna posebna obuka za korisnike kako bi naučili koristiti novi korisnički interfejs. Drugi problem su formati datoteka, jer se može desiti da ne koriste svi najnoviju verziju aplikacije, pa se može desiti da dokument kreiran novom verzijom ne mogu otvoriti korisnici koji još koriste staru verziju.

Kako su operativni sistemi svakodnevno izloženi napadima zlonamjernim software-om, važno je obezbijediti redovno instaliranje "zakrpa" (slika 5). Korisnicima je ostavljena mogućnost da se ta instalacija vrši automatski, u određeno vrijeme, ili da sami odrede kad i šta treba ažurirati i instalirati. Kod izbora vremena, treba voditi računa da ta instalacija usporava rad sistema, opterećuje mrežnu konekciju, ali i da često zahtijeva restart sistema, što može dovesti do smetnji u radu ili čak gubitka podataka.

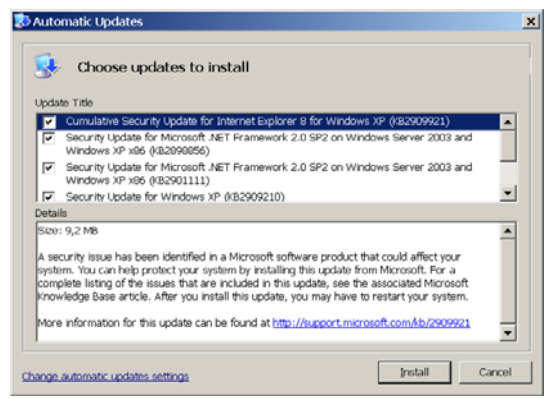

*Slika 5. Instalacija "zakrpa" za Windows XP*

### 4.4. Zaštita od zlonamjernog software-a

Zlonamjerni software se najčešće uprošteno naziva "računarskim virusima", ali on obuhvata daleko širi dijapazon prijetnji. Računarski virusi se danas najčešće prenose putem prenosivih medija kao što su USB *flash drive*, ali se ove i druge vrste zlonamjernog software-a mogu prenijeti i na druge načine: pokretanjem skripti na kompromitovanim web stranicama, otvaranjem dokumenata koji sadrže zlonamjerni kôd u vidu skripti, itd. Korisnici često nesvjesno instaliraju zlonamjerni software tako što instaliraju besplatne pomoćne alate (npr. *ASK toolbar* za navodno brže pretraživanje interneta vrši promjene na računaru korisnika koje je teško kontrolisati i koji smanjuju sigurnost).

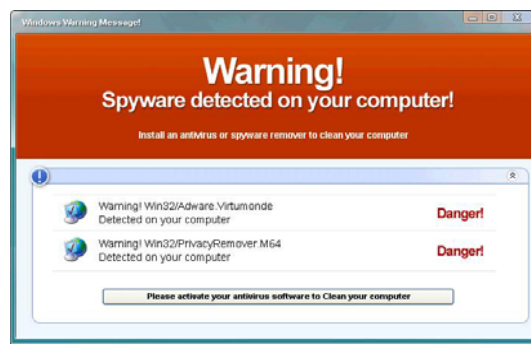

Naivni i neiskusni korisnici često nasjedaju na lažne poruke o antivirusnim programima, koji koriste poruke i interfejse koji samo podsjećaju na pravi antivirusni software, a u stvari čine štetu računaru (slika 6).

*Slika 6. Lažni antivirusni software*

## 5. ODRŽAVANJE KOMUNIKACIJSKE OPREME
### 5.1. Lokalne mreže

Pravilno projektovanje lokalne mreže, uključujući korištenje struktuiranog kabliranja, neprekidnih izvora napajanja i planiranje kapaciteta, značajno smanjuje potrebe za održavanjem. Za razliku od korisničke opreme (personalnih računara), održavanje ovog dijela računarske opreme se rijetko prepušta korisnicima, nego se za to obično angažuju specijalizirani davatelji tih usluga ili isporučioci mrežne opreme.

Održavanje lokalne mreže koje radi administrator sistema se svodi na vođenje evidencije o IP adresama, imenima računara u mreži, pravima korisnika, praćenje neovlaštenih upada u sistem i zloupotreba sistema, te dijeljenju resursa (štampači i prostor za pohranu podataka). Za neometan rad sistema je veoma važno uspostaviti i ažurirati evidenciju tih podataka. U ovaj dio održavanja spada i održavanje lokalnog servera, ukoliko on postoji.

## 5.2. Internet konekcija

Za kvalitet i funkcionisanje internet konekcije bi trebalo da garantuje davatelj usluga (*internet provider*). U nekim slučajevima odgovornost ipak može biti i na krajnjem korisniku. primjer za to je pregrijavanje modema ili druge komunikacijske opreme, koje može dovesti do smetnji u vezama i pada propusnosti konekcije. Drugi primjer mogućeg uzroka usporenja internet konekcije je loše podešen sistem ažuriranja sistema (svi korisnici istovremeno vrše ažuriranje operativnog sistema ili antivirusnog software-a). Samo jedan računar zaražen zlonamjernim software-om može opteretiti konekciju tako da blokira sve ostale korisnike.

Mrežni administratori često blokiraju pristup nekim internet servisima, kako bi rasteretili internet konekciju. Jednostavnim dodavanjem IP adrese *127.0.0.1* u datoteci *hosts.txt* za adrese kao što su *Youtube.com*, *Facebook.com*, *Rapidshare.com* i slične servise koji nisu neophodni za obavljanje redovnog posla blokira se pristup korisnicima i bez dodatne mrežne opreme.

## 5.3. Bežične komunikacije

Sve raširenija mogućnost bežičnog pristupa lokalnim mrežama u organizacijama olakšava poslovanje, jer svaki korisnik na prenosnom računaru, tabletu ili mobilnom telefonu ima pristup mrežnim resursima. S druge strane, otvara se novi problem sigurnosti, jer se time stvara mogućnost da bilo ko u neposrednoj blizini može dobiti pristup lokalnoj mreži. Zato je vrlo važno uspostaviti sistem za sigurnost konekcije, što može biti politika periodične promjene šifre za pristup, korištenje fiksnih IP adresa ili neki drugi način zaštite.

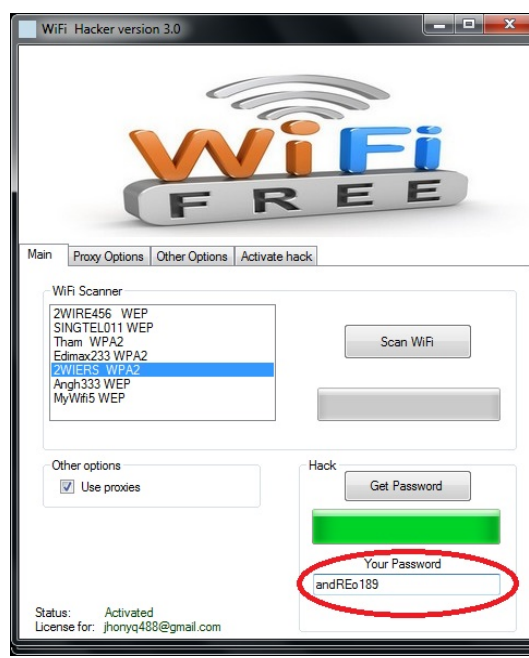

*Slika 7. Software za otkrivanje šifri za pristup bežičnoj mreži*

## 6. PRISTUPI ODRŽAVANJU

Pristup održavanju u najvećoj mjeri zavisi od složenosti organizacije. Mali poslovni subjekti, kao što su obrtničke radnje, sa 2-3 računara ne mogu priuštiti vlastitog administratora sistema i za to koriste usluge vanjskog dobavljača. Izbor strategije održavanja zavisi prvenstveno od troškova, zatim od važnosti brzine reagovanja, ali je vrlo važan i aspekt sigurnosti, odnosno povjerljivosti onoga koji vrši održavanje. nekad se kombinuju različiti pristupi, tako da se za dio poslova održavanja zaduže i obuče sami korisnici, dok se za složenije zadatke koriste specijalizirane usluge vanjskog dobavljača.

## 6.1. Vlastiti nespecijalizirani kadrovi

Administrator sistema mora posjedovati širok spektar informatičkih kompetencija, a takav kadar je na visokoj cijeni. Ne može svaka organizacija priuštiti diplomiranog inženjera elektrotehnike ili informatike sa više godina iskustva, specijalističkim certifikatima (CCNA, CCNP, MS) i potrebnim vještinama. Zato se često jedan od postojećih zaposlenih koji ima najviše sklonosti ka informacionim tehnologijama  opredijeli i zaduži da vrši te poslove. Takav pristup jeste naizgled najjeftiniji i ima najbrži odziv, ali može se pokazati kao neprihvatljiv u nekim situacijama, jer nedostatak stručnosti može dovesti do nemjerljivih posljedica.

## 6.2. Vlastita specijalizirana IT služba

Najkvalitetnije rješenje je uspostavljanje vlastite specijalizirane službe unutar organizacije, koja pruža informatičku podršku svim korisnicima. Kod definisanja opisa poslova, potrebnih kvalifikacija i resursa za takvu službu, treba voditi računa da oni obuhvate sve ranije pomenute aspekte, od hardware-a, software-a, komunikacija, do sigurnosti i razvoja informacionog sistema. ta služba treba biti zadužena za evidencije, planiranje troškova za IT resurse, obuku kadrova, kao i kontakte s dobavljačima opreme i software-a.

Ukoliko organizacija ima implementiran sistem upravljanja kvalitetom, lakše će definisati uloge i zadatke ove službe.

## 6.3. Korištenje vanjskih usluga

Ako se organizacija opredijeli da poslove održavanja računarske opreme povjeri vanjskom dobavljaču, bilo da se radi o angažmanu administratora sistema po ugovoru o djelu ili da koristi usluge pravnog lica koje se bavi održavanjem, najvažnije je obezbijediti povjerljivost, odnosno sigurnost informacija. Prvi korak u tom smislu je potpisivanje ugovora o povjerljivosti, kojim će se dobavljač obavezati da sve informacije do kojih dođe ne smije koristiti izvan organizacije, te da se predvide rigorozne i provedive sankcije ukoliko do toga ipak dođe.

Drugi veoma važan kriterij kod izbora vanjskog dobavljača usluga održavanja računarske opreme je brzina odziva. U zavisnosti od prirode posla organizacije, nekad je neophodno obezbijediti pravovremeno reagovanje. Razlikuju se ti kriteriji za bankarski sektor, obrazovanje, prodaju, proizvodnju itd. Nekad čak i zakonski propisi ograničavaju vrijeme odziva, na primjer izdavanje fiskalnih računa podrazumijeva da sve komponente funkcionišu besprijekorno: od računara, komunikacijske linije, do štampača.

U praksi su prisutna dva modela troškova održavanja: fiksni trošak podrazumijeva paušalno plaćanje koje podrazumijeva obaveze dobavljača da reaguje i otkloni kvarove u zadatom roku, dok tarifiranje po pozivu obično podrazumijeva definisanu cijenu sata angažmana servisera. U oba slučaja važno je jasno definisati ko ima pravo angažovati usluge održavanja, ko daje odobrenje i ko procjenjuje je li zaista utrošeno onoliko vremena koliko je dobavljač fakturisao naručiocu.

Posebno zanimljiv pristup održavanju je *outsourcing*, odnosno prenošenje dijela funkcija organizacije na vanjske dobavljače. Primjer takvog pristupa je virtualizacija, pomoću koje se usluge masovne pohrane podataka umjesto na vlastitom serveru realizuju na dijeljenim resursima provajdera. Tako se automatski eliminišu troškovi održavanja (administratorsko osoblje, licence za sistemski software, *backup*, održavanje hardware-a, i sl. Takav pristup u BiH koriste uglavnom veliki poslovni sistemi, iako bi se mogle ostvariti značajne uštede optimalnom primjenom tog pristupa i u malim i srednjim preduzećima.

## 6.4. Dokumentacija

Dokumentacija računarskih sistema obuhvata ne samo liste osnovnih sredstava, nego i evidencije korisnika, mrežnih adresa, licenci software-a, kao i evidencije korištenja (*logs*). Na slici 8 prikazana je evidencija korištenja personalnog računara, iz koje se može dijagnosticirati software-ski kvar. Na slici 9 dat je primjer evidencije pokušaja prijavljivanja korisnika na server, iz kojeg se može uočiti neovlašteni pristup sistemu.

Važan dio dokumentacije održavanja predstavlja plan održavanja. S obzirom da održavanje računarske opreme podrazumijeva brojne aspekte, od kojih su samo neki ovdje nabrojani, od velikog značaja je pravilno planiranje. Plan održavanja treba kreirati na osnovu evidencije opreme i software-a, a treba da sadrži informacije o tome šta treba uraditi, kada i ko je odgovoran za realizaciju pojedinog zadatka.

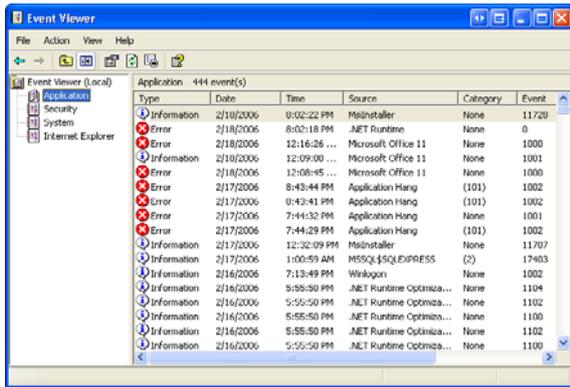

*Slika 8. Praćenje evidencije grešaka u radu operativnog sistema*

*Slika 9. Evidencija svih pokušaja pristupa korisnika serveru*

### 6.5. Važnost permanentnog obrazovanja

Brzina razvoja i promjena informacionih tehnologija diktira obavezu stalnog usavršavanja osoblja, kako korisnika, tako i administratora sistema. Svaka promjena verzije software-a, konfiguracije sistema ili pojava novih alata i servisa podrazumijeva period prilagođavanja koji mora biti pokriven obukom. Dodatna obuka mora biti sastavni dio plana održavanja i nadgradnje računarskih sistema.

### 7. ZAKLJUČAK

Održavanje računarskih sistema je složen zadatak, kako za krajnje korisnike, tako i za menadžment svake organizacije. Od veličine organizacije, stepena informatizacije i od karaktera poslovanja zavisi opredjeljenje načina održavanja. Bez obzira na izabrani i implementirani način, neophodno je uspostaviti i voditi urednu dokumentaciju, koja troškove održavanja može smanjiti na minimum. Prvi korak u uspostavljanju sistema održavanja je detaljna evidencija raspoloživog hardware-a i software-a, na osnovu koje se vrši planiranje.